\documentclass[preprint,10pt]{elsarticle}

\usepackage{graphicx, booktabs, tabularx, hyperref, listings, rotating}
\usepackage{amsmath}
\usepackage{cleveref}
\usepackage{natbib}

\title{Benchmarking 34 OpenKIM Nickel Potentials with an Emphasis on Surfaces and Extended Defects}
\author[1]{Matthew Thoms}
\author[1,2]{Hao Sun}
\author[1]{Laurent Karim B\'{e}land}
\affiliation[1]{organization={Department of Mechanical and Materials Engineering, Queen's University},
addressline={130 Stuart Street},
city={Kingston},
postcode={K7L 3N6},
country={Canada}}
\affiliation[2]{organization={Liaoning Academy of Materials},
addressline={109-3 Quanyun Road},
city={Shenyang},
postcode={110167},
country={China}}
\date{\today}

\begin{document}

\begin{abstract}
We present an automated benchmarking suite for face-centered-cubic (FCC) nickel that evaluates 47 quantitative metrics spanning both standard tests (equation of state, elastic constants, surface energies and phonons) and application-specific scenarios such as defect formation and migration, grain boundaries, step edges, close-range interactions, and vacancy cluster energetics. Using this framework, we assess 34 interatomic potentials from the OpenKIM repository, including pairwise, embedded-atom, modified-embedded-atom, angular-dependent, and spectral neighbor analysis potentials (SNAP). Results are compared against \textit{ab initio} benchmarks compiled from the literature. Most potentials accurately reproduce lattice parameters, elastic constants, and surface energies, whereas predictive accuracy degrades for migration barriers and short-range compression. Principal-component analysis identifies correlated property groups and a partially orthogonal component associated with migration and short-range physics, revealing Pareto trade-offs between accuracy domains. SNAP models occupy the lowest-error frontier, although several embedded-atom potentials remain competitive across many metrics. The framework provides a reproducible baseline for potential selection, highlights systematic limitations across formalisms, and supports benchmarking-in-the-loop strategies for developing next-generation machine-learning potentials for Ni and Ni-based alloys.
\end{abstract}

\maketitle

\begin{sidewaystable}[ph!]
    \centering
    \begin{tabularx}{1.1\textwidth}{l X l X}
        \toprule
        \# & Model & \# & Model \\
        \midrule
        1 & \detokenize{EAM_CubicNaturalSpline_AngeloMoodyBaskes_1995_Ni} & 19 & \detokenize{EAM_Dynamo_ZhouJohnsonWadley_2004_Ni} \\
        2 & \detokenize{EAM_Dynamo_AcklandTichyVitek_1987_Ni} & 20 & \detokenize{EAM_Dynamo_ZhouJohnsonWadley_2004NISTretabulation_Ni} \\
        3 & \detokenize{EAM_Dynamo_AcklandTichyVitek_1987v2_Ni} & 21 & \detokenize{EAM_IMD_BrommerGaehler_2006A_AlNiCo} \\
        4 & \detokenize{EAM_Dynamo_AdamsFoilesWolfer_1989Universal6_Ni} & 22 & \detokenize{EAM_IMD_BrommerGaehler_2006B_AlNiCo} \\
        5 & \detokenize{EAM_Dynamo_BonnyCastinTerentyev_2013_FeNiCr} & 23 & \detokenize{EMT_Asap_Standard_JacobsenStoltzeNorskov_1996_AlAgAuCuNiPdPt} \\
        6 & \detokenize{EAM_Dynamo_BonnyPasianotCastin_2009_FeCuNi} & 24 & \detokenize{Morse_Shifted_GirifalcoWeizer_1959HighCutoff_Ni} \\
        7 & \detokenize{EAM_Dynamo_Foiles_1985_Ni} & 25 & \detokenize{Morse_Shifted_GirifalcoWeizer_1959LowCutoff_Ni} \\
        8 & \detokenize{EAM_Dynamo_FoilesBaskesDaw_1986Universal3_Ni} & 26 & \detokenize{Morse_Shifted_GirifalcoWeizer_1959MedCutoff_Ni} \\
        9 & \detokenize{EAM_Dynamo_MendelevBorovikov_2020_FeNiCr} & 27 & \detokenize{Sim_LAMMPS_ADP_MishinMehlPapaconstantopoulos_2005_Ni} \\
        10 & \detokenize{EAM_Dynamo_MendelevKramerHao_2012_Ni} & 28 & \detokenize{Sim_LAMMPS_MEAM_AsadiZaeemNouranian_2015_Ni} \\
        11 & \detokenize{EAM_Dynamo_Mishin_2004_NiAl} & 29 & \detokenize{Sim_LAMMPS_MEAM_EtesamiAsadi_2018_Ni} \\
        12 & \detokenize{EAM_Dynamo_MishinFarkasMehl_1999_Ni} & 30 & \detokenize{Sim_LAMMPS_MEAM_KoGrabowskiNeugebauer_2015_NiTi} \\
        13 & \detokenize{EAM_Dynamo_MishinMehlPapaconstantopoulos_2002_NiAl} & 31 & \detokenize{Sim_LAMMPS_MEAM_Wagner_2007_Ni} \\
        14 & \detokenize{EAM_Dynamo_OnatDurukanoglu_2014_CuNi} & 32 & \detokenize{SNAP_LiHuChen_2018_Ni} \\
        15 & \detokenize{EAM_Dynamo_PunMishin_2009_NiAl} & 33 & \detokenize{SNAP_LiHuChen_2018_NiMo} \\
        16 & \detokenize{EAM_Dynamo_TehranchiCurtin_2010_NiH} & 34 & \detokenize{SNAP_ZuoChenLi_2019_Ni} \\
        17 & \detokenize{EAM_Dynamo_WilsonMendelev_2015_NiZr} & 35 & \detokenize{SNAP_ZuoChenLi_2019quadratic_Ni} \\
        18 & \detokenize{EAM_Dynamo_ZhangAshcraftMendelev_2016_NiNb} & & \\
        \bottomrule
    \end{tabularx}
    \caption{List of tested interatomic potentials and their corresponding indices used in subsequent figures.}
    \label{tab:potential key}
\end{sidewaystable}

\begin{sidewaystable}[ph!]
	\centering
	\begin{tabularx}{\textwidth}{X X}
		\toprule
		\textbf{Common} & \textbf{Less Common} \\
		\midrule
		Elastic constants & Adsorbed defect formation/migration\\
		Equation of state & Bulk point defect migration\\
		Free surface formation & Melting point\\
		Grain boundary formation & Quasi-static drag\\
		$\gamma$-Surface (GSFE) & Step edge formation\\
		Phonon density-of-states & Vacancy cluster coalescence\\
		Bulk point defect formation & \\
		\bottomrule
	\end{tabularx}
	\caption{Summary of common and less common benchmark tests available in major interatomic potential repositories.}
	\label{tab:test breakdown}
\end{sidewaystable}

\section{Introduction}

Materials behavior reflects processes operating across many scales, from electronic bonding to macroscopic deformation.
Atomistic simulations help connect these scales by resolving how defects form, migrate, and interact at nanometer and picosecond scales that are inaccessible to experiments.
The accuracy of such simulations depends directly on the quality of the interatomic model used.
Compared with \textit{ab initio} methods, interatomic potentials make it possible to simulate much larger systems over longer times, though with some loss of fidelity and transferability beyond the configurations for which they were parameterized.

Because of this trade-off, it has become standard practice to record and compare the material properties predicted by different potentials.
Shared databases such as the NIST Interatomic Potential Repository (IPR) and the Open Knowledgebase of Interatomic Models (OpenKIM) compile results for a broad range of single- and multi-element systems. \cite{BECKER2013277,tadmor:elliott:2011}
These repositories focus on well-defined “common” benchmarks—equilibrium volumes, elastic constants, surface energies, generalized stacking-fault energies, phonon spectra, and point-defect formation energies—which serve as a first filter when selecting a potential for a given study.
Such comparisons, together with direct validation against higher-accuracy calculations, help establish whether the efficiency of an interatomic potential is a reasonable compromise for the intended application.

Many scenarios that matter in practical applications are still not well covered by standard benchmarks.
Examples include surface and bulk migration barriers, complex interfaces such as grain boundaries and step edges, short-range forces relevant to collision cascades or ablation, and the energetics of vacancy clusters.
Understanding how different potential formalisms perform across this wider set of tests is important when selecting, calibrating, or training models for systems exposed to corrosion, irradiation, or high-temperature service.
Benchmarking coverage also depends strongly on the material system: each element or alloy must be evaluated separately, with attention to the crystallographic structures and defect types that are most relevant.
While the NIST IPR allows automated property retrieval through tools such as \texttt{atomman} and \texttt{iprpy}, \cite{Hale_2018} OpenKIM still relies on potential authors to submit individual results, which makes broad, consistent benchmarking difficult.

Because of the large number of possible tests even for a single element, a complementary strategy is to analyze how groups of related properties vary together.
Techniques such as principal component analysis (PCA) make this possible by reducing the dimensionality of the dataset and identifying clusters of correlated metrics.
Instead of considering each benchmark in isolation, one can examine the collective behavior of metrics that respond similarly across different models.
This view helps reveal systematic trends and trade-offs—for instance, between structural and dynamical properties—and can guide the refinement of fitting strategies or the design of new potential formalisms.

Nickel provides a useful test case for comprehensive benchmarking.
Chemically, nickel ions play central roles in organometallic catalysts used in chemical synthesis and in compounds relevant to hydrogen fuel cell technologies.
Structurally, nickel and its alloys are indispensable in high-temperature environments, from jet turbines to steam generators in nuclear power plants, because they retain mechanical stability at large fractions of their melting point while resisting corrosion.
These applications involve highly dynamic processes, and understanding them at the macroscopic level depends on accurate atomistic descriptions of defect formation, migration, and interaction at surfaces and interfaces.
Assessing how current interatomic models represent elemental nickel therefore provides a foundation for developing reliable models of more complex and technologically relevant systems.

In this work we develop an automated molecular dynamics benchmarking suite for FCC systems that evaluates 47 metrics, with particular attention to defect-related phenomena such as point and surface defects, grain boundaries, and vacancy clusters.
These properties are rarely included in existing repositories. The suite is fully documented and publicly available, allowing other researchers to reproduce the tests, extend them to new materials, or compare additional potentials.
We apply this framework to 34 Ni–Ni potentials available in the OpenKIM database and compare their predictions with density functional theory (DFT) benchmarks from the literature.
Using principal component analysis, we identify correlations among metrics and evaluate accuracy trade-offs across different potential formalisms.

The remainder of this article is organized as follows.
Section 2 introduces the potential formalisms and summarizes their main features.
Section 3 describes the benchmarking methodology and the specific tests implemented.
Section 4 presents and discusses the results, including the PCA and defect energetics.
Section 5 concludes with a summary of the main findings and perspectives for future model development.

\section{Methods}

\subsection{Potential Details}

We tested 42 Ni--Ni potentials from the OpenKIM database, of which 35 unique models are analyzed here after removing duplicates (see Table \ref{tab:potential key}). These models fall into four broad categories: pairwise potentials, embedded-atom method (EAM) potentials, derivatives of the EAM formalism, and machine-learning potentials. One exception is Potential~24, which belongs to the effective medium theory (EMT) family developed by Jacobsen \emph{et~al.} \cite{OpenKIM-MO:115316750986:001a}

Pairwise potentials approximate the total energy of a system as the sum of interactions between all atomic pairs. They balance a short-range repulsion with a long-range attraction, producing a stable equilibrium bond length. The most familiar example is the Lennard--Jones potential. In this study we have included potentials adopting the Morse formalism, defined as \cite{PhysRev.34.57}

\begin{align}
    U_{\text{tot}} &= \tfrac{1}{2}\sum_i \sum_{j \ne i} U(r_{ij}), \\
    U(r_{ij}) &= D_e \left[ 1 - \exp\!\left( -a (r_{ij} - r_e) \right) \right]^2 ,
\end{align}

where $D_e$ is the well depth, $r_e$ is the equilibrium bond length, and $a$ controls the width of the potential well, with larger $a$ producing a narrower well. With only three adjustable parameters, the Morse formalism cannot capture more subtle bonding features, but it is simple, broadly transferable, and computationally efficient for large-scale simulations. The three Morse potentials tested here were parameterized to reproduce the experimental lattice constant, sublimation energy, and compressibility of nickel at 0~K and zero pressure.

The embedded-atom method (EAM), introduced by Daw and Baskes, extends pairwise potentials by adding an ``embedding'' term that correlates the energy of each atom with the local electron density generated by its neighbors \cite{DAW1993251}. The total energy is written as

\begin{equation}
    U_{\text{tot}} = \sum_i G_i \!\left( \sum_{j \ne i} \rho_j(r_{ij}) \right)
    + \tfrac{1}{2}\sum_i \sum_{j \ne i} U(r_{ij}) ,
\end{equation}

where $\rho_j(r_{ij})$ represents the local electron density contribution from atom $j$, and $G_i$ is the embedding energy of atom $i$ in this density. This approach provides a simplified description of the electron gas similar in spirit to density functional theory. Because the formalism depends on metallic electron densities, EAM potentials are best suited for metals rather than for materials with strongly directional covalent bonding.

In practice, the embedding function $G_i$ and density function $\rho_j$ are usually tabulated, as in the Sandia DYNAMO format, with interpolation performed using cubic Hermite splines. Potentials~2--21 in Table~\ref{tab:potential key} follow this tabulation scheme, representing nearly 60\% of the unique models tested. Potential~1 is a minor variation that employs cubic natural splines for interpolation. The EAM formalism remains widely used in molecular dynamics and Monte Carlo simulations of metals because it combines computational efficiency, achieved through tabulated functions, with enough flexibility to reproduce a broad range of material properties.

A limitation of the EAM formalism is that it depends only on radial functions of interatomic separation, which limits its ability to describe directional bonding. To address this, several variants have been proposed that include angular terms. The most widely used is the modified embedded-atom method (MEAM), which introduces a screening factor $S_{ij}$ into the pairwise interaction term \cite{PhysRevB.46.2727}. This screening depends on the geometry of atom triplets and attenuates interactions that are blocked by intervening atoms. In doing so, MEAM can represent directional bonding and stabilize a wider range of crystal structures than the basic EAM formalism.

The angular-dependent potential (ADP) formalism, developed by Mishin, Mehl, and Papaconstantopoulos for the Fe--Ni system, is another extension of EAM that improves the treatment of directionality \cite{MISHIN20054029}. Unlike MEAM, which introduces angular effects indirectly through screening functions, ADP includes them explicitly by adding dipole and quadrupole distortion terms to the energy expression. The pairwise interaction, electron density, and distortion tensor contributions are all given by analytic functions with adjustable parameters. For a single-element system, ADP requires 20 fitted parameters, increasing to 29 for the Fe--Ni binary alloy. While ADP is less flexible than a fully tabulated EAM potential, it encodes more physical information and offers specific advantages for modeling binary alloy systems.

The effective medium theory (EMT) potential developed by Jacobsen \emph{et~al.} was designed as a simple and transferable model for face-centered-cubic (FCC) metals, and as a foundation for more general ``universal'' potential formalisms. \cite{OpenKIM-MO:115316750986:001a} The total energy is written as

\begin{equation}
    U_{\text{tot}} = \sum_i U_{c,i}(n_i) + \Delta U_{\text{AS}} + \Delta U_{\text{1el}},
\end{equation}

where $U_{c,i}(n_i)$ is a cohesive function of the local electron density $n_i$, $\Delta U_{\text{AS}}$ is an atomic-sphere correction, and $\Delta U_{\text{1el}}$ is a one-electron correction. The latter two terms are defined relative to a reference system that is similar to, but not identical with, the target metal, and they are mapped onto the system through pair potentials. In total, EMT uses only seven fitted parameters. Jacobsen \emph{et~al.} emphasized that EMT is approximate by design: although informed by \emph{ab initio} data, it sacrifices some absolute accuracy in favor of simplicity and transferability across different metals.

In recent years, machine-learning interatomic potentials have emerged as an important evolution of classical fitted models. These methods span a wide range of formalisms, from tensor-contraction approaches with strong physical grounding to more flexible neural-network architectures. Their common goal is to combine the scalability of interatomic potentials with the near-quantum accuracy of \emph{ab initio} calculations. In this study, Potentials~31--34 correspond to spectral neighbor analysis potentials (SNAPs), which use bispectrum descriptors of the local atomic environment to construct linear or quadratic regression models of the potential energy surface.

Table~\ref{tab:potential roots} summarizes the tested potentials, grouped by the number of system components and their constituent elements. Among all formalisms, EAM potentials represent the largest share of the sample. SNAP models form the next most numerous group and include the greatest number of potentials developed specifically for elemental nickel rather than for alloyed systems. All EAM potentials for pure nickel were benchmarked against self-diffusion data, with Potential~12 fitted directly to \emph{ab initio} references. Potential~16 follows this approach for the Ni--H system.

\begin{figure}
	\centering
	\includegraphics[width=.5\textwidth]{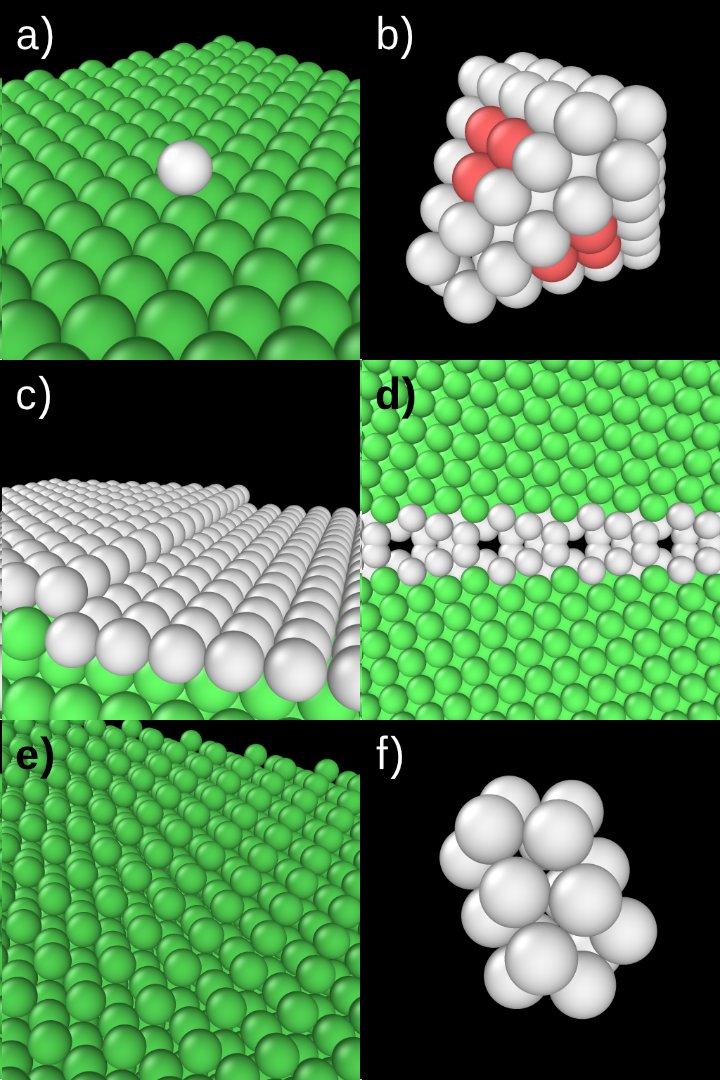}
	\caption{Representative atomic configurations used in the benchmarking suite:  
(a) adatom on a (111) surface,  
(b) stacking-fault tetrahedron formed by ten vacancies,  
(c) (100)--(111) step edge,  
(d) $\Sigma9$ tilt grain boundary,  
(e) (332) free surface, and  
(f) six-vacancy void.}

	\label{fig:OVITO renders}
\end{figure}

\begin{table*}
    \centering
    \begin{tabularx}{\textwidth}{X X X X X}
        \toprule
         \textbf{Formalism} & \textbf{Components} & \textbf{System} & \textbf{Potentials} & \textbf{Notes}\\
         \midrule
         EAM & 1 & Ni & 2 & \\
          & & & 3 & \\
          & & & 4 & \\
          & & & 8 & \\
          & & & 12 & \\
          & 2 & Ni-H & 1 & Cubic spline interpolation \\
          & & & 16 & \\
          & & Ni-Cu & 7 & \\
          & & & 14 & \\
          & & Ni-Zr & 10 & \\
          & & & 17 & \\
          & & Ni-Al & 11 & \\
          & & & 13 & \\
          & & & 15 & \\
          & & Ni-Nb & 18 & \\
          & 3 & Ni-Fe-Cr & 5 & \\
          & & & 9 & \\
          & & Ni-Fe-Cu & 6 & \\
          & & Ni-Fe-Co & 19 & \\
          & & & 20 & Retabulation of 19 \\
          & & Ni-Al-Co & 21 & \\
          & & & 22 & Retabulation of 21 \\
         \midrule
         EMT & 7 & Al-Ag-Au-Cu-Ni-Pd-Pt & 23 & Universal for system components\\
         \midrule
         Morse & 1 & Ni & 24 & \\
         & & & 25 & \\
         & & & 26 & \\
         \midrule
         ADP & 2 & Fe-Ni & 27 & \\
         \midrule
         MEAM & 1 & Ni & 29 & \\
          & & & 31 & \\
          & 2 & Ni-Ti & 30 & \\
          & 3 & Ni-Al-Cu & 28 & \\
         \midrule
         SNAP & 1 & Ni & 32 & \\
          & & & 34 & \\
          & & & 35 & Quadratic model \\
          & 2 & Ni-Mo & 33 & \\
         \bottomrule
    \end{tabularx}
\caption{Summary of the tested interatomic potentials grouped by formalism, number of components, and material system. Numerical indices correspond to those listed in Table~\ref{tab:potential key}.}

    \label{tab:potential roots}
\end{table*}

\begin{figure*}
    \centering
    \includegraphics[width=0.5\textwidth]{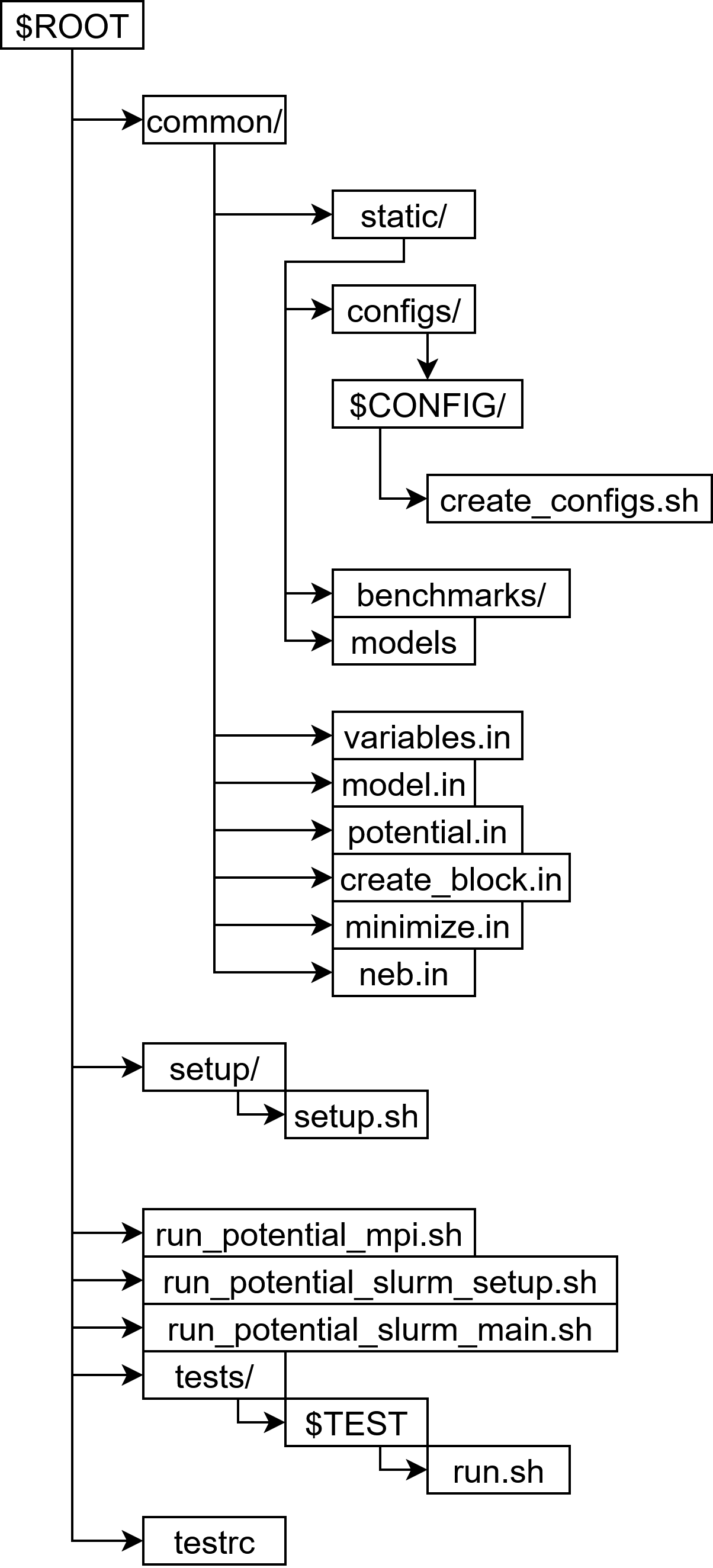}
\caption{Overall directory structure of the benchmarking suite. Folder names prefixed with the symbol~`\$' represent groups of folders that serve the same function for different tests.}

    \label{fig:dir structure}
\end{figure*}

\begin{figure*}[t!]
	\centering
	\includegraphics[width=\textwidth]{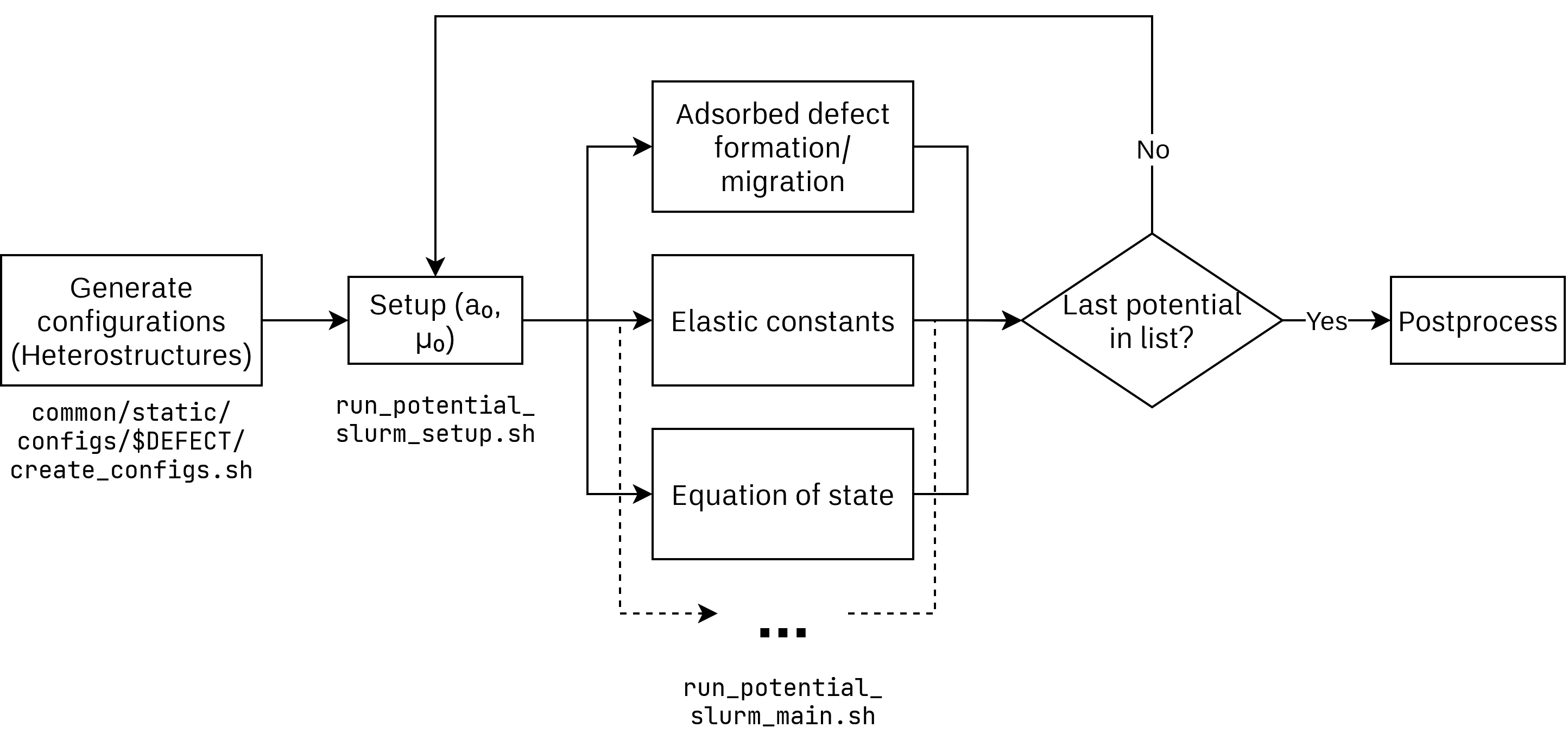}
\caption{Workflow of the benchmarking suite on a high-performance computing (HPC) system using a job scheduler. The initial setup job determines the equilibrium lattice parameter and per-atom reference energy. Subsequent tests are then submitted and executed in parallel to minimize total runtime.}
	\label{fig:flowchart}
\end{figure*}

\subsection{Test Details}

\begin{figure*}[t!]
    \centering
    \makebox[\textwidth][c]{%
        \includegraphics[width=1.5\textwidth]{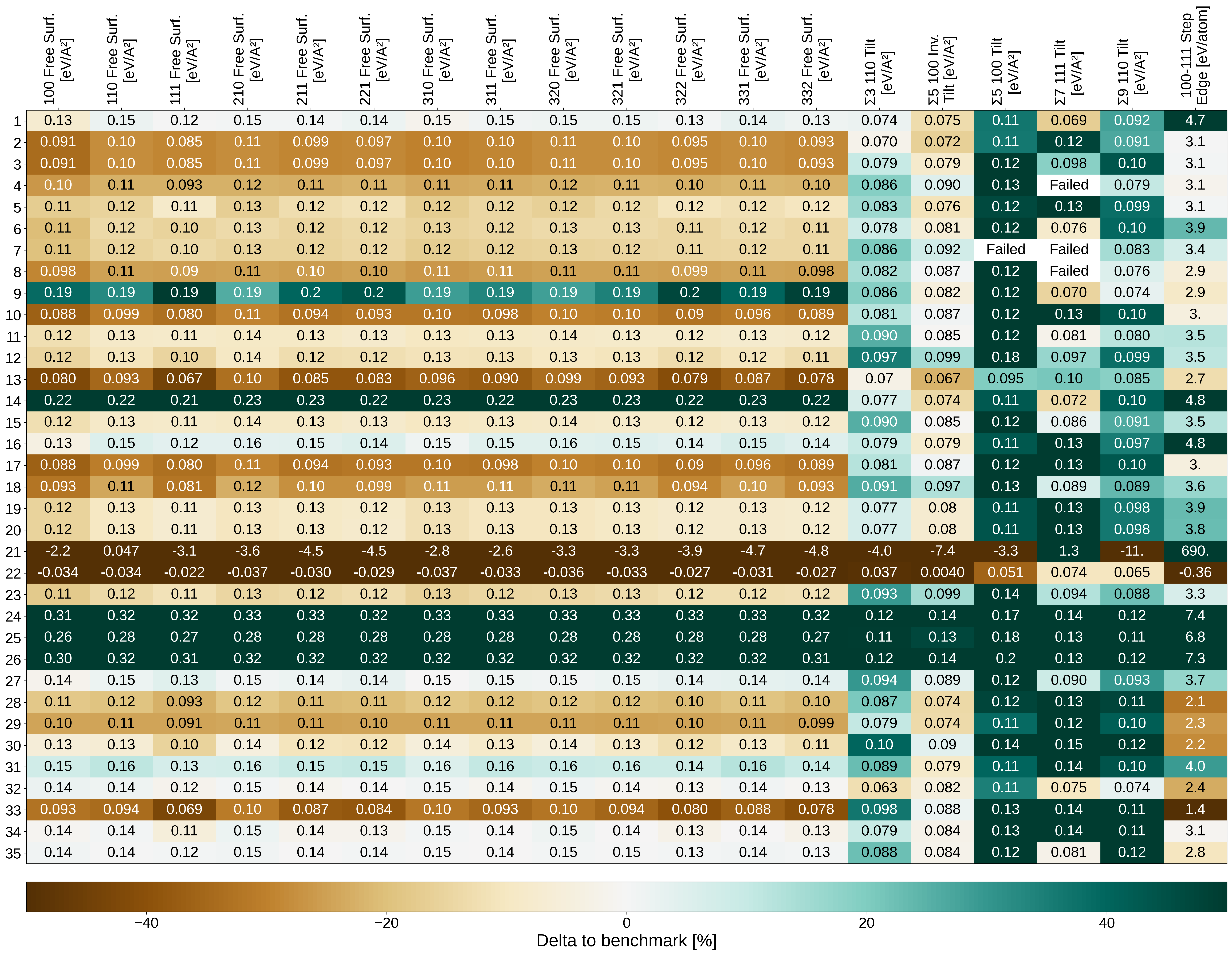}%
    }
    \caption{Comparison of interatomic potentials with DFT reference values for free-surface, grain-boundary, and step-edge formation energies. 
    Colors indicate the relative deviation from benchmark values (\%). 
    Horizontal striping in the free-surface data highlights strong intra-potential correlations: 
    a potential that reproduces one surface energy accurately tends to reproduce others with similar accuracy.}
    \label{fig:large heterostruct}
\end{figure*}

\begin{figure*}[t!]
	\centering
	\makebox[\textwidth][c]{%
        \includegraphics[width=1.3\textwidth]{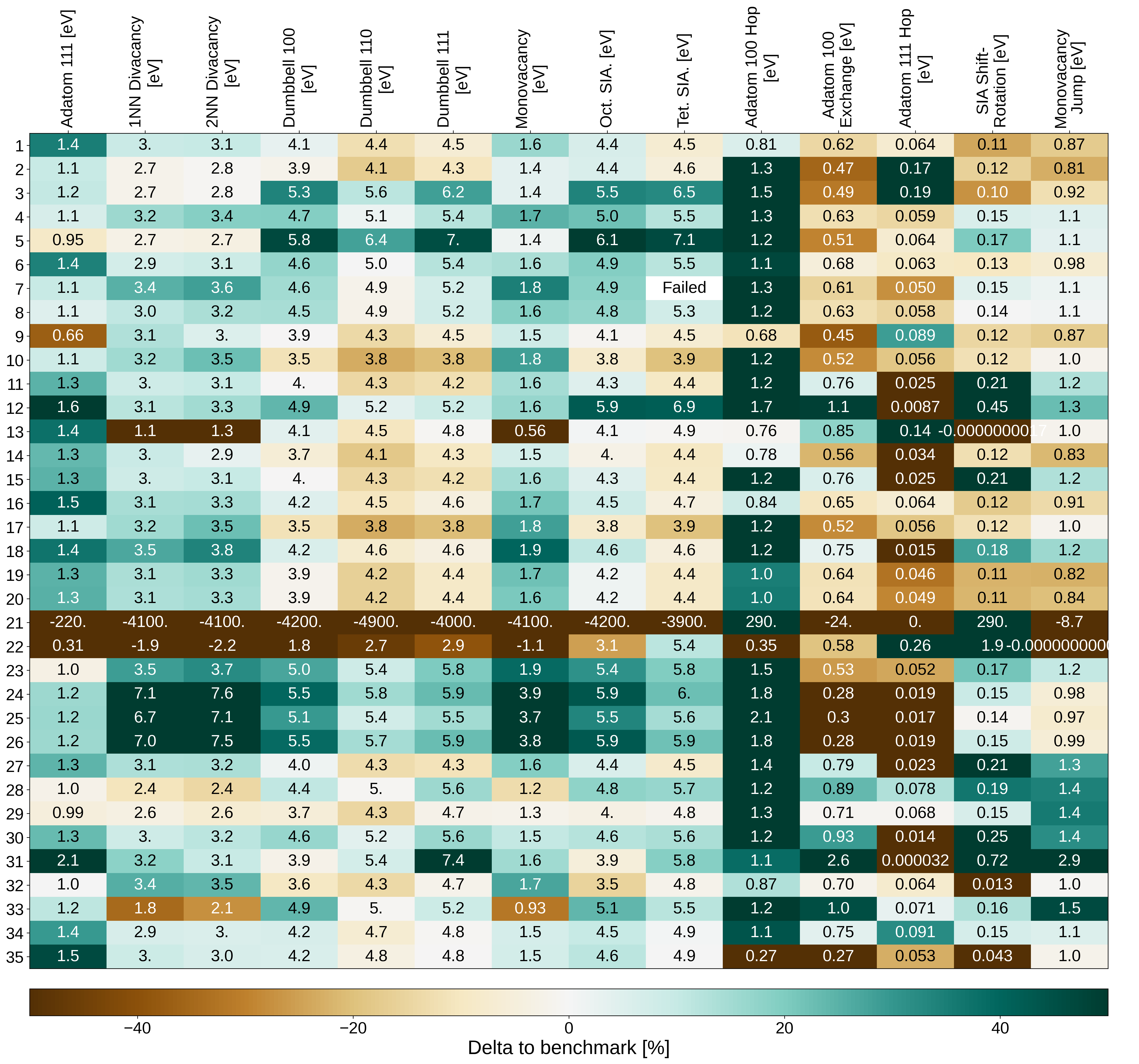}%
    }
\caption{Comparison of interatomic potentials with DFT reference values for the formation and migration energies of bulk and surface point defects. 
Colors indicate the relative deviation from the benchmark values (\%). 
Unlike the surface-energy benchmarks, intra-potential correlations are weak, resulting in noisier patterns across metrics. 
Migration energies, particularly for surface defects, are poorly reproduced by most potentials.}

	\label{fig:point defect}
\end{figure*}

\begin{figure*}[t!]
	\centering
	\makebox[\textwidth][c]{%
        \includegraphics[width=1.3\textwidth]{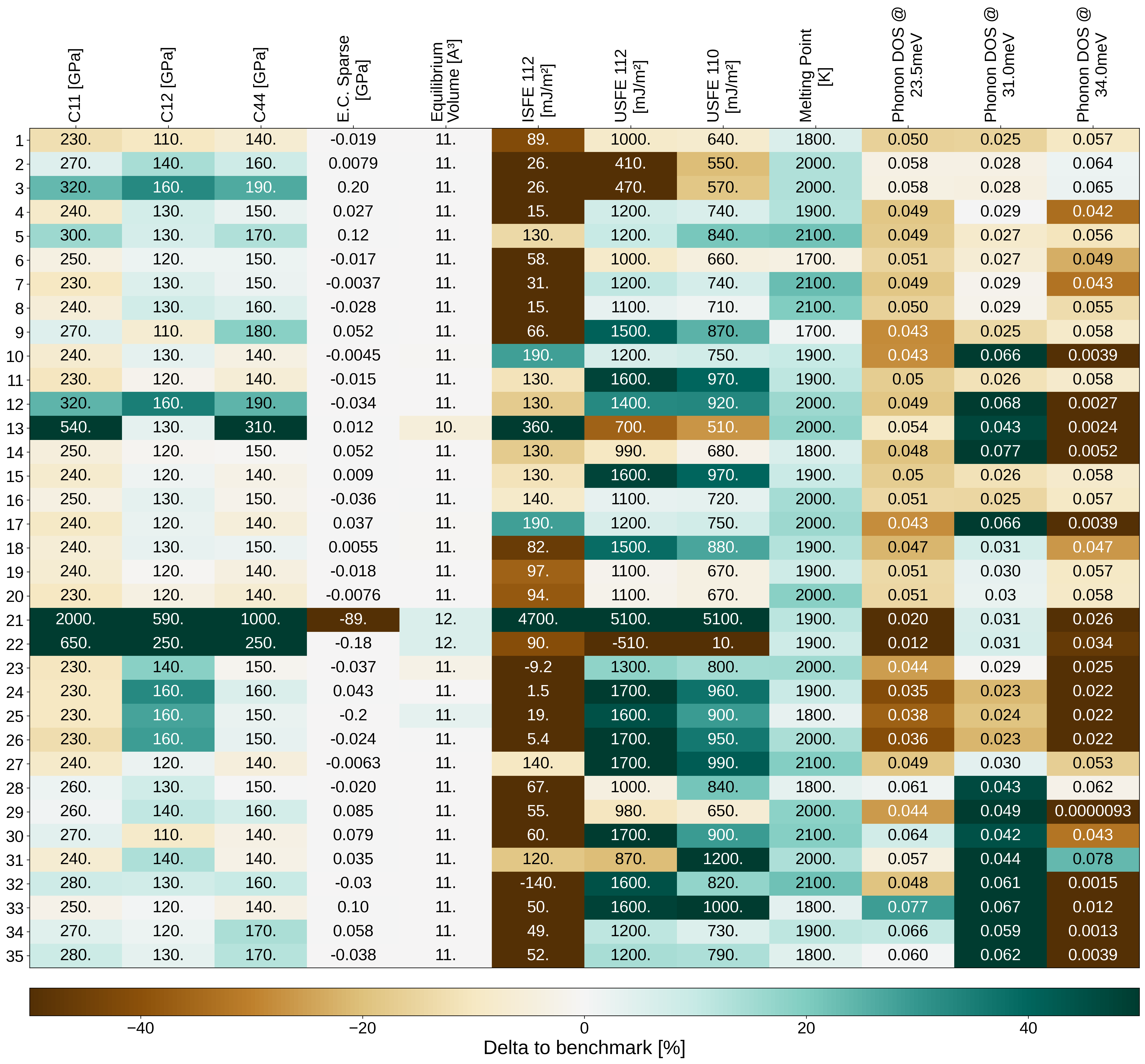}%
    }
\caption{Comparison of calculated material constants and other relevant energies for all tested interatomic potentials. Colors indicate the relative deviation from the benchmark values (\%). 
Overall agreement with benchmarks is good, although specific intra-model trends remain difficult to identify.}

	\label{fig:mtl constants}
\end{figure*}

\begin{figure*}[t!]
	\centering
	\includegraphics[width=.5\textwidth]{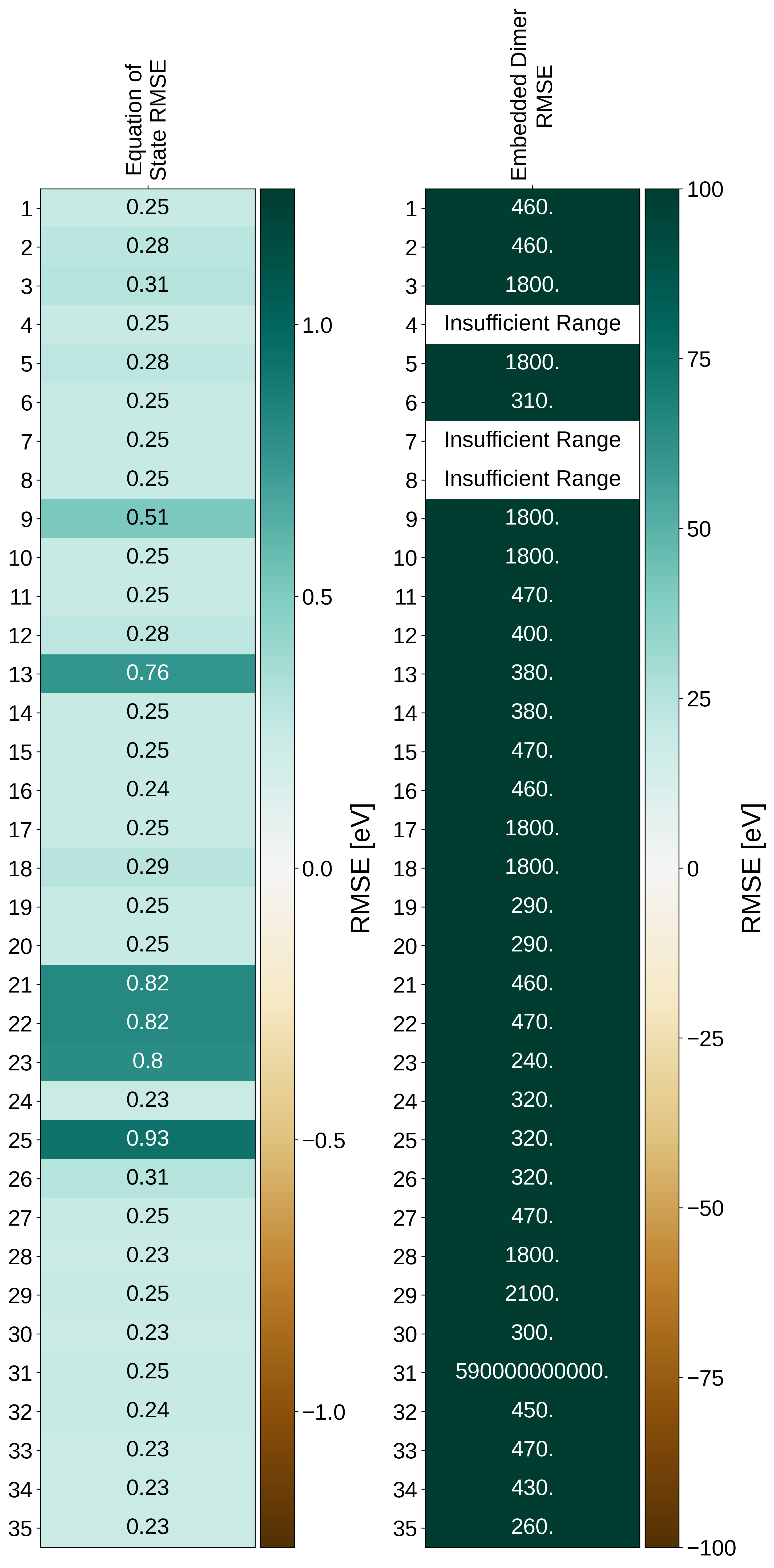}
\caption{Root mean square error (RMSE) of curve-based metrics, including the generalized stacking-fault energy (GSFE), equation of state, phonon density of states (DOS), and quasi-static drag.  The left panel shows the absolute RMSE values, while the right panel indicates range-based flags (frequency range out of bounds for phonons and insufficiently short radius parameterization for QSD).  RMSE values are generally homogeneous across potentials, although the deviations from the benchmark curves differ in character.}

	\label{fig:RMSE}
\end{figure*}

\begin{figure*}[t!]
	\centering
	\includegraphics[width=.5\textwidth]{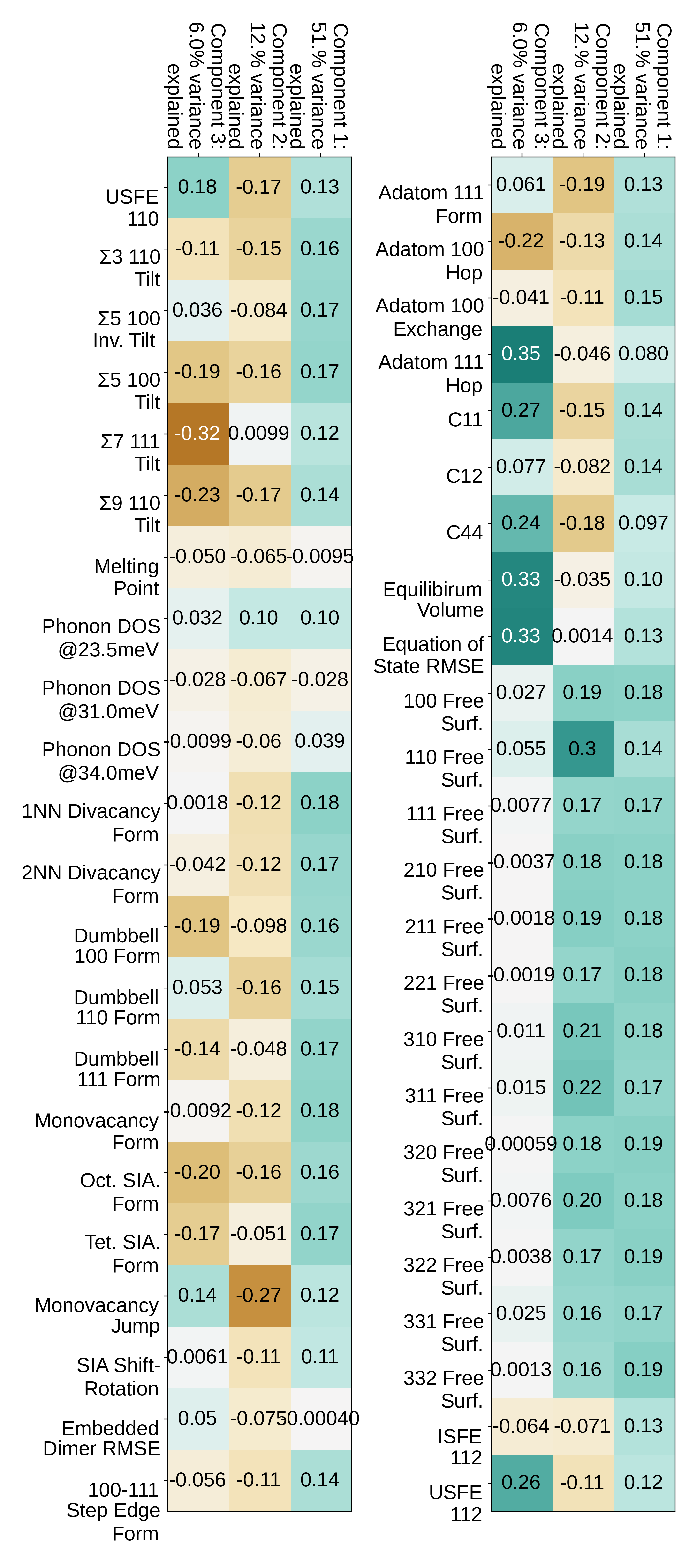}
\caption{PCA weight maps for the first three components. 
Blue shading corresponds to positive weights, and orange shading to negative weights.  Together, the three components explain nearly 70\% of the total variance in the dataset. Component~3 shows a degree of orthogonality with the equation-of-state metrics and point-defect diffusion, whereas many other metrics are correlated within the same component.}

	\label{fig:pca weights}
\end{figure*}

\begin{figure*}[t!]
	\centering
	\makebox[\textwidth][c]{%
        \includegraphics[width=1.2\textwidth]{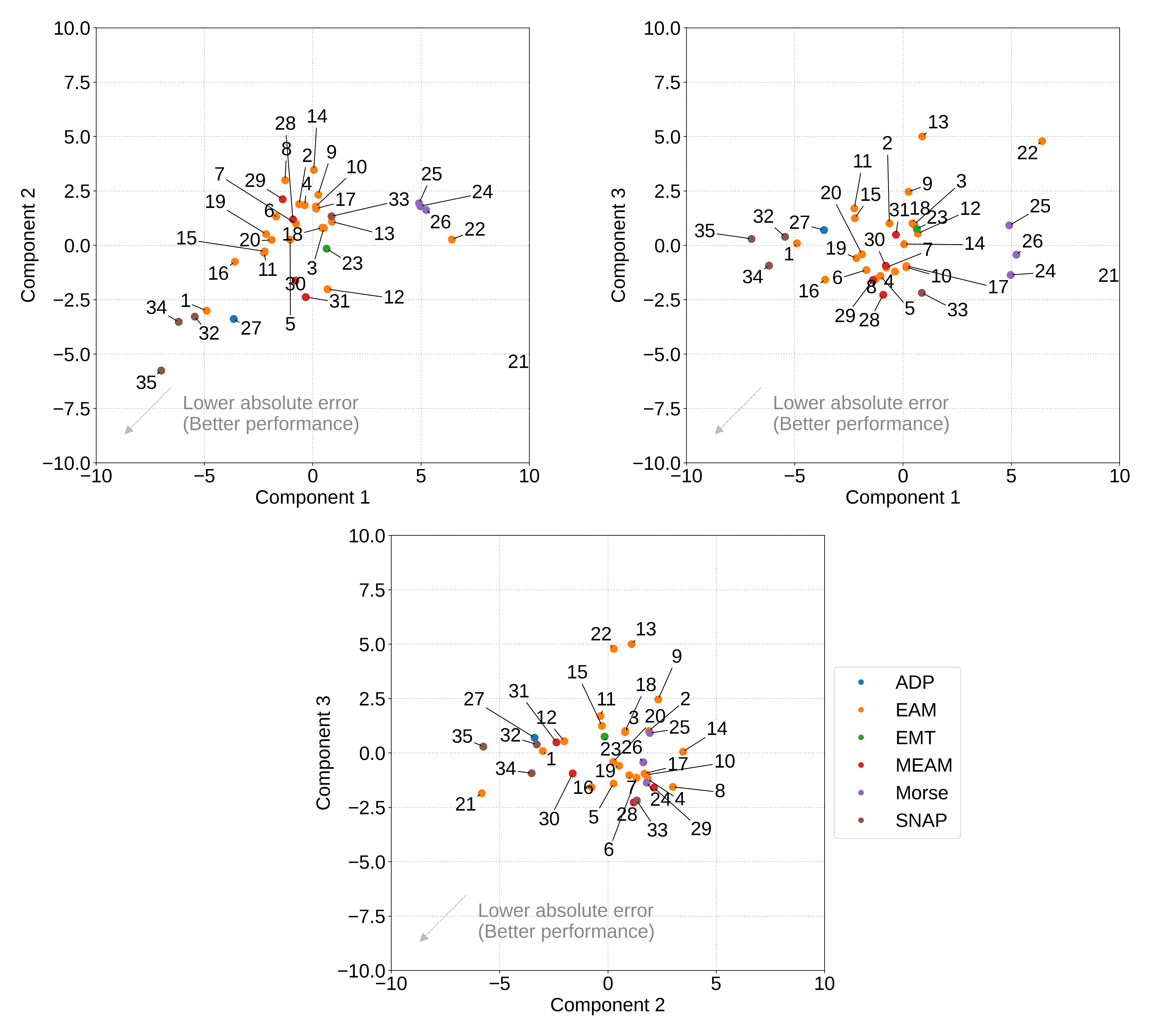}%
    }
\caption{Pairwise plots of the first three principal PCA components based on log-normalized absolute error. Scatterplot markers correspond to the index of each potential.  Comparisons between the two non-orthogonal components reveal strong positive correlations, with SNAP potentials occupying the lowest-error region. When comparing with the orthogonal component, emerging Pareto fronts become apparent, particularly between Components~1 and~3.}

	\label{fig:pca plots}
\end{figure*}

\begin{figure*}[t!]
	\centering
	\includegraphics[width=.7\textwidth]{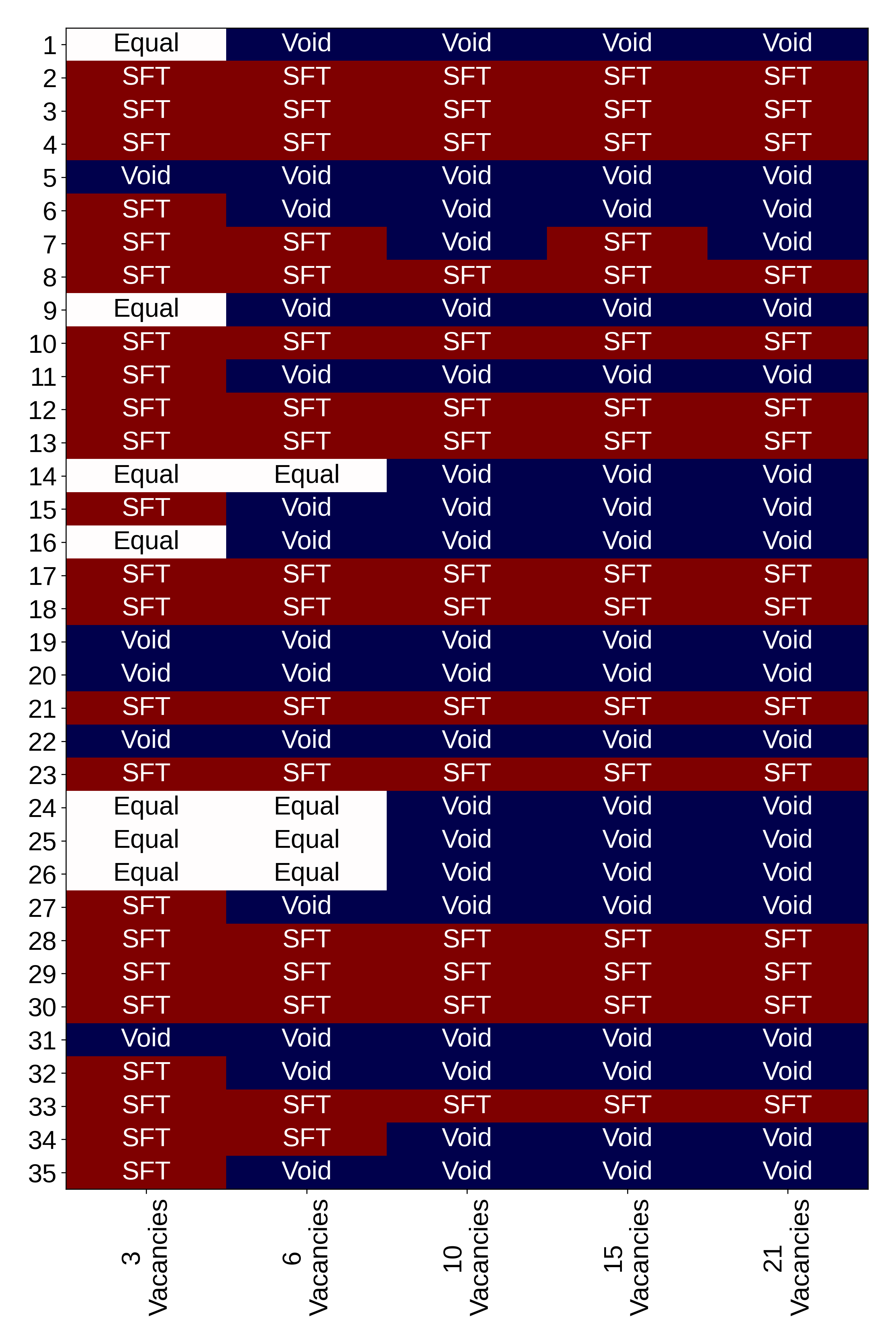}
\caption{Formation energies of vacancy clusters, including stacking-fault tetrahedra (SFTs) and amorphous voids, for clusters containing 3, 6, 10, 15, and 21 vacancies. SFTs are generally favored at small sizes (three vacancies), whereas voids become more stable at larger sizes for most potentials.  A few potentials, however, favor SFTs across all cluster sizes.}

	\label{fig:vacancy clust}
\end{figure*}

All simulations were performed using LAMMPS \cite{THOMPSON2022108171}, with interatomic potentials loaded through the OpenKIM API \cite{elliott:tadmor:2011}.  Simulations employed metal units and a timestep of 1~fs. Unless otherwise noted, energy minimizations used convergence tolerances of $1\times10^{-10}$~eV for the total energy and $1\times10^{-10}$~eV{\AA}$^{-1}$ for forces, with a maximum of 5{,}000 steps and 50{,}000 energy or force evaluations. Standard damping constants were 0.1~ps for Nose--Hoover thermostats and 1~ps for Nose--Hoover barostats. 
Periodic boundary conditions were applied along all three Cartesian directions unless otherwise specified. The complete list of metrics evaluated in this study is provided in Table~\ref{tab:test breakdown}.

The first ``setup'' test performed for each potential determines the equilibrium face-centered-cubic (FCC) lattice parameter and chemical potential. The lattice parameter $a_0$ was obtained by scanning the instantaneous pressure of a four-atom unit cell over lattice parameters ranging from 2.5~\AA\ to 4.5~\AA\ in 0.01~\AA\ increments and identifying the zero-pressure point by root finding. The chemical potential $\mu_0$ was defined as the per-atom energy after minimization at this equilibrium lattice parameter, given by $\mu_0 = U_{\text{tot}}/4$, where $U_{\text{tot}}$ is the total potential energy of the system. This procedure avoids reliance on the LAMMPS \texttt{fix box/relax} algorithm, which can produce false equilibria, and yields results consistent with those reported in the NIST Interatomic Potential Repository. Because many subsequent tests require these equilibrium values to define initial configurations and reference energies, this setup stage is executed first so that all other calculations can proceed in parallel without redundant computation.

The equilibrium volume and equation-of-state data were calculated using the same approach as for the equilibrium lattice parameter. 
A four-atom cell was scanned over lattice parameters ranging from 0.8 to 8.0~\AA\ in 0.01~\AA\ increments, and the potential energy per atom was recorded without minimization.
RMSE accuracy is only calculated over the range of 3.45 to 3.59~AA to match the benchmark from Prandini \emph{et~al.} \cite{prandini_2021_69mvv-cv520}
The resulting energy–lattice-spacing curve was interpolated with cubic splines, and the equilibrium point was identified by minimizing the spline fit using the BFGS algorithm.

Short-range repulsion was probed by displacing a single atom toward its first and second nearest neighbors within a 2048-atom cell, from 0 to \(0.95\,a_0\) in increments of \(0.001\,a_0\), while recording the instantaneous energy. 
Potentials that were not parameterized for this interatomic distance range were flagged as out of range. 
This procedure is sometimes referred to as the quasi-static drag test; here, it is denoted as the embedded-dimer test.

Elastic constants were calculated at 300~K using the LAMMPS \texttt{compute born/matrix} command.
A 108--atom cell was simulated under canonical (NVT) conditions with a Nose--Hoover thermostat for 200~ps.

Simulation configurations for all defect- and interface-related calculations, including defect formation and migration energies, grain-boundary and step-edge formation energies, free-surface formation energies, generalized stacking-fault energy (GSFE) curves, and vacancy-coalescence tests, were pre-generated using Atomsk \cite{HIREL2015212} and then relaxed in LAMMPS.
All configurations were initially built at a lattice parameter of 3.52~\AA\ and subsequently rescaled to the calculated equilibrium FCC lattice parameter $a_0$.

Free surfaces (Figure~\ref{fig:OVITO renders}e) were generated in Atomsk by rotating an orthogonally aligned FCC unit cell so that the surface normal aligned with the $z$-axis.
The minimum periodic cell was recalculated during this transformation, and the system was replicated along $z$ until the total height exceeded 50~\AA.
The box size was then doubled in the $z$-direction to create a vacuum region. The free-surface formation energy after minimization was computed as

\begin{equation}
    U_{\text{surf}} = \frac{U_{\text{tot}} - n\mu_0}{2A},
\end{equation}

where $U_{\text{tot}}$ is the total potential energy, $n$ is the number of atoms, and $A$ is the surface area in the $x$--$y$ plane. 
The factor of $\tfrac{1}{2}$ accounts for the creation of two surfaces (top and bottom) in the simulation cell.

Grain boundaries (Figure~\ref{fig:OVITO renders}d) were constructed in a similar way.
Two crystal blocks were created by tilting a bulk region by half the misorientation angle on each side of the grain-boundary plane so that the boundary normal lay along the $z$-axis.
The simulation box was extended along $z$ to at least 100~\AA\ to minimize self-interaction.
The grain-boundary formation energy was calculated using the same expression as for free surfaces, with the area $A$ corresponding to the $x$--$y$ plane.

Step edges (Figure~\ref{fig:OVITO renders}c) were generated in a manner similar to the free-surface configurations, but with the bulk of atoms replicated multiple times along the $x$ and $y$ axes (16~$\times$~16 for the (100)--(110) and (111)--(100) terraces, and 20~$\times$~16 for the (100)--(111) terrace). 
After constructing a floating bulk analogous to a free-surface cell, a one-atom-thick strip was removed from one of the surfaces over half of the total box length in the $x$-direction. 
Before computing the step-edge formation energy, the free-surface formation energy for each system was recalculated as described above. 
The step-edge formation energy was then obtained after minimization as

\begin{equation}
    U_{\text{step}} = \frac{U_{\text{tot}} - n\mu_0 - 2U_{\text{surf}}A_{\text{surf}}}{2},
\end{equation}

where $A_{\text{surf}}$ is the surface area in the $x$--$y$ plane.

\medskip

Point defects were divided into two categories: bulk and surface. 
Bulk defects were created in 6912-atom cells using Atomsk by adding, removing, or displacing atoms as required. 
Their formation energies were computed after minimization as

\begin{equation}
    U_{\text{defect}} = U_{\text{tot}} - n\mu_0.
\end{equation}

Surface point defects (“addefects”) were generated on 4000-atom floating bulks similar to those used for the free-surface tests. 
Adatoms (Figure~\ref{fig:OVITO renders}a) were placed at lattice sites on the (100) and (111) surfaces. 
For (111) adatoms, the atom was displaced 0.25~\AA\ toward the surface to prevent delamination during relaxation in LAMMPS. 
Advacancies were formed by removing atoms at specified sites. 
The corresponding formation energies were computed after minimization as

\begin{equation}
    U_{\text{defect}} = U_{\text{tot}} - n\mu_0 - 2U_{\text{surf}}A_{\text{surf}},
\end{equation}

where $A_{\text{surf}}$ again denotes the surface area in the $x$--$y$ plane.

\medskip

Migration barriers for all defects were determined using the same pre-generated configurations as their formation-energy counterparts. 
Atoms were displaced into their final states to create initial and final configurations, which were then minimized. 
The nudged elastic band (NEB) method was applied using the LAMMPS \texttt{fix neb} and \texttt{neb} commands to obtain both forward and reverse migration barriers. 
A force tolerance of 0.01~eV~\AA$^{-1}$ and a limit of 5000 steps were used for both standard and climbing-image NEB runs.

\medskip

Vacancy-cluster structures (Figures~\ref{fig:OVITO renders}b and~\ref{fig:OVITO renders}f) were evaluated using the same approach as for bulk point defects. 
Defects were introduced in a 16{,}384-atom supercell generated with Atomsk, then rescaled and minimized in LAMMPS. 
Formation energies were calculated using the same expression as for bulk point defects.

Generalized stacking-fault energy (GSFE) calculations were performed using a 288-atom system composed of two slabs rotated such that the (111) plane was aligned with the $z$-axis. 
Each slab was approximately 146.3~\AA\ thick. 
To eliminate artificial symmetries in the resulting $\gamma$-surface, the boundary condition along $z$ was set to \texttt{shrinkwrap} rather than periodic. 
The system was first minimized with the slabs perfectly aligned to record the reference energy $U_{\text{base}}$. 
The upper slab was then shifted incrementally along both Burgers vectors, $\vec{b}$, where the [112] direction was aligned with $x$ and the [110] direction with $y$. 
Displacements ranged from $0.0\vec{b}$ to $1.0\vec{b}$ in steps of $0.001\vec{b}$. 
During each minimization (including the initial configuration), atoms were allowed to relax only along $z$. 
The GSFE was calculated as

\begin{equation}
    U_{\text{GSFE}} = \frac{U_{\text{tot}} - U_{\text{base}}}{A},
\end{equation}

where $U_{\text{tot}}$ is the minimized total energy and $A$ is the interfacial area in the $x$--$y$ plane.

\medskip

Melting points were determined from phase-coexistence simulations using a 512-atom system containing 256 atoms in both the solid and liquid regions. 
Simulations were performed at temperatures between 1250 and 2500~K in 5~K increments. 
The coexistence structure was prepared by heating half of the simulation cell to 3000~K under canonical (NVT) conditions while keeping the other half fixed, followed by relaxation to remove atomic overlap. 
The combined system was then equilibrated under isothermal–isobaric (NPT) conditions at zero pressure, allowing the solid–liquid interface to move freely. 
To extract the melting point, the quarter-maximum broadening of the first peak in the radial distribution function (RDF) was evaluated for each temperature. 
This quantity produces a sigmoid curve with a characteristic rise near $T_m$. 
The curve was smoothed with a 21-point rolling average, and linear fits were applied to the lower (1250–1400~K) and upper (2350–2500~K) regions. 
The minimum distance between these fitted lines was computed across the entire temperature range, and a Gaussian was fit to this distribution. 
The mean of the Gaussian provided the estimated melting temperature.

\medskip

The harmonic phonon density of states (DOS) was calculated using \textsc{Phonopy} \cite{Togo_2023,doi:10.7566/JPSJ.92.012001} interfaced with LAMMPS. 
A sampling grid of \(101\times101\times101\) $k$-points and a Gaussian broadening of 0.2~meV were used, matching the DFT+$U$ benchmark of Bhardwaj \emph{et~al.} \cite{bhardwaj2023latticedynamicsrelatedproperties}.

\subsection{Analysis Details}

Reference data from density functional theory (DFT) were collected from a wide range of literature sources. 
It is important to note that these reference values were obtained using diverse computational settings, including different pseudopotentials, exchange–correlation functionals, supercell sizes, energy cutoffs, and $k$-space sampling densities. The corresponding parameters are reported in Appendix \ref{appendix:benchmark}. As a result, the DFT benchmarks are not strictly uniform, but rather represent typical high-accuracy values that can be used to assess whether a given interatomic potential offers an acceptable trade-off between accuracy and computational cost relative to force-field molecular dynamics.

Principal component analysis (PCA) was used to identify correlations among benchmark metrics. Absolute errors from each potential were treated as samples, and all data were log-normalized prior to analysis. 
Missing values, which occurred primarily in tests probing short-range interactions where certain models were evaluated outside their fitted range, were imputed using the mean value of the corresponding metric across all potentials. Three orthogonal components were retained, as additional components contributed less than 1\% to the total explained variance. 
Each potential’s performance along the three components was then compared through pairwise plots. Because of the log normalization, lower mean errors correspond to more negative coordinates in the PCA space. All PCA computations were carried out using the \texttt{scikit-learn} Python package. \cite{scikit-learn}

\subsection{Usage Details}

The benchmarking suite was designed with an emphasis on portability, automation, modularity, and efficient parallel execution. 
All configuration and automation were implemented using shell scripts. 
The only mandatory dependencies for running the tests were a POSIX-compliant shell environment, a shared-library build of LAMMPS with the KIM package, Python, and \textsc{Phonopy}. 
Configuration pre-generation was handled separately from testing using Python and \textsc{Atomsk}, the latter serving as an optional dependency for users wishing to modify or regenerate initial configurations.

An overview of the project directory structure is shown in Figure~\ref{fig:dir structure}. 
Common sections of the LAMMPS input scripts were modularized and stored in the \texttt{./common} directory. 
Each test script begins by including \texttt{./common/variables.in} and \texttt{./common/model.in}, which define standard variables and import the selected potential from the OpenKIM database using metal units. 
The default parameters in \texttt{./common/variables.in} can be overridden by redefining them earlier in the script. 
Once the system geometry is defined, the script includes \texttt{./common/potential.in} to set interaction parameters, thermodynamic output intervals, and neighbor-list update settings. 
Other modules in this directory implement reusable operations: \texttt{./common/create\_block.in} constructs a four-atom FCC unit cell at a specified lattice parameter, \texttt{./common/minimize.in} performs an energy minimization with standard criteria, and \texttt{./common/neb.in} carries out a nudged elastic band (NEB) calculation. 
The \texttt{./common/static} directory contains static resources that are not executed during LAMMPS runs, including pre-generated configurations, benchmark datasets for post-processing, and the list of OpenKIM models to be evaluated.

All global configuration settings were defined in the \texttt{./testrc} file, where users specified paths to executables, MPI and scheduler settings, and environment modules for systems using Lmod. 
The list of tested potentials was stored as KIM identifiers in \texttt{./common/static/models}. 
Data for each potential were collected independently in both workstation and HPC environments. 
On a local system, after installing the dependencies, users executed \texttt{./run\_potential\_mpi.sh} to perform all tests (including the setup step) sequentially for a single potential; this process was repeated for each model as needed. 
On high-performance computing systems, the workflow followed the structure shown in Figure~\ref{fig:flowchart}. 
After verifying dependencies, users executed \texttt{./run\_potential\_slurm\_setup.sh} to submit the setup job (calculation of $a_0$ and $\mu_0$) to a SLURM scheduler. 
Once this job completed, \texttt{./run\_potential\_slurm\_main.sh} submitted all remaining tests to run in parallel. 
These two steps were repeated for each potential until the full benchmark suite was completed.

\section{Results and Discussion}

\subsection{Individual Metrics}

Figures~\ref{fig:large heterostruct}--\ref{fig:RMSE} summarize the performance of the 35 unique Ni--Ni potentials across all evaluated metrics relative to DFT benchmarks. 
Each figure presents tests along the $x$-axis and potentials along the $y$-axis, ordered as in Table~\ref{tab:potential key}. 
Figures~\ref{fig:large heterostruct}--\ref{fig:mtl constants} display single-value metrics as percentage deviations from benchmark values, while Figure~\ref{fig:RMSE} reports root-mean-square errors (RMSE) for continuous curves such as the equation of state, generalized stacking fault energy (GSFE), phonon density of states (DOS), and quasi-static drag. 
In all plots, white indicates minimal error, whereas darker orange or green corresponds to larger deviations. 
Entries marked as ``Failed'' denote tests that did not converge after three attempts, typically due to inadequate short-range parameterization preventing relaxation.

The arrangement of related metrics in the heatmaps highlights covariance between tests. 
Horizontal stripes of similar color indicate consistent performance of a potential across related metrics, whereas vertical stripes reveal similar behavior among many potentials for a given test. 
Random or noisy patterns suggest little or no correlation between metrics.

\medskip
\noindent\textbf{Surface and interface formation energies.}  
Figure~\ref{fig:large heterostruct} shows results for free-surface, grain-boundary, and step-edge formation energies. 
Free-surface energies are strongly correlated across all orientations: if a potential reproduces one surface energy well, it tends to reproduce others with similar accuracy. 
Most potentials underestimate surface energies relative to DFT benchmarks, with only a few (notably the Morse and two EAM models) showing moderate overestimation. 
Several EAM potentials systematically overestimate surface formation energies by 10–30\%. 
In contrast, more flexible formalisms such as ADP, MEAM, and SNAP achieve higher accuracy overall.

Grain-boundary formation energies exhibit higher variance and weaker correlations. 
They are generally overestimated, especially for the $\Sigma 5$ (100) tilt boundary, which forms a vertical stripe across the heatmap—indicating that many models struggle to fully relax this configuration. 
For the $\Sigma 9$ (110) boundary, EAM potentials occasionally outperform more advanced formalisms such as ADP, MEAM, or SNAP. 
However, only EAM potentials failed to complete some tests due to insufficient short-range coverage. 
Step-edge formation energies show partial inverse correlation with free-surface results: several potentials that perform well for surfaces (e.g., Potentials~1 and~16) overestimate step-edge energies.

\medskip
\noindent\textbf{Defect formation and migration energies.}  
Figure~\ref{fig:point defect} presents formation and migration energies for bulk and surface point defects. 
These data are significantly noisier, indicating weak intra-potential correlation. 
Formation energies are generally accurate, even for weaker models such as the Morse potentials, though adsorbed-atom (adatom) configurations are less well reproduced. 
Tetrahedral self-interstitials are consistently more difficult to form than octahedral ones, and formation energies tend to be slightly overestimated. 

Migration barriers show larger discrepancies. 
Surface defect migration is particularly challenging: adatom migration on the (100) surface is systematically overestimated, while other surface defects are underestimated. 
Potentials that accurately capture surface energies also tend to reproduce (100) adatom migration barriers well. 
For bulk defects, EAM models scatter more randomly around the benchmarks, whereas ADP, MEAM, and SNAP potentials systematically overestimate migration energies.

\medskip
\noindent\textbf{Elastic and thermodynamic properties.}  
Figure~\ref{fig:mtl constants} summarizes elastic constants, equilibrium lattice parameters, and other bulk properties, with corresponding RMSEs shown in Figure~\ref{fig:RMSE}. 
Elastic constants show good agreement with DFT, exhibiting low overall noise. 
Some systematic biases are observed: the Morse potentials overestimate $C_{12}$, while EAM potential~13 overestimates both $C_{11}$ and $C_{44}$. 
Equilibrium lattice parameters are well reproduced across all models. 
Equation-of-state RMSEs cluster around 0.25~eV, and many potentials reproduce melting behavior within $\sim$200~K of the benchmark, though the results are generally biased toward overestimation. 
MEAM potentials yield the largest overestimations of melting points, followed by EAM potentials~5, 7, and~8. 

For GSFE and phonon DOS curves, many models qualitatively match the DFT profiles but differ at key points. 
Most underestimate the intrinsic stacking-fault energy while overestimating the unstable stacking-fault energy, which may affect the predicted mobility of partial dislocations. 
SNAP potentials in particular show deviations in the phonon DOS, notably in the 31–34~meV range, where several peaks appear flattened compared with DFT benchmarks. 
Similar behavior occurs in some MEAM and EAM models.

\medskip
\noindent\textbf{Short-range behavior.}  
At short interatomic distances, many models deviate strongly from DFT references, as shown in Figure~\ref{fig:RMSE}. 
This is expected, since none of the tested potentials were explicitly fitted for collision cascades or high-energy interactions, and no hybridization with ZBL-type potentials was performed. 
The MEAM potential of Wagner \emph{et~al.} exhibits especially steep repulsion, while some models fail to relax entirely in this regime and are labeled as ``Insufficient Range.'' 
Such cases reappear in subsequent defect tests where short-range parameterization is again required.

\medskip
\noindent\textbf{Overall trends.}  
Across all tests, the weakest performers are the Morse potentials (24–26) and the Brommer--Gähler 2006 IMD models (21–22), both of which yield unphysical energies for several metrics. 
The latter discrepancies likely stem from successive reparameterizations for incompatible simulation frameworks rather than intrinsic flaws in the functional form.

\subsection{Principal Component Analysis}

Figure~\ref{fig:pca weights} summarizes the weights associated with the three retained principal components. 
Green cells correspond to positive contributions and yellow cells to negative ones. 
Together, these three components account for approximately 70\% of the total variance in the dataset; additional components contribute less than 1\% each and were therefore omitted.

\medskip
\noindent\textbf{Component interpretation.}  
Component~1 captures the broadest range of metrics, with notable weighting from free-surface, grain-boundary, and point-defect formation energies. 
Component~2 is dominated by positive weighting from free-surface energies, counterbalanced by negative contributions from $\gamma$-surface metrics, elastic constants, and several defect-formation and migration energies. 
Component~3 emphasizes elastic constants, equation-of-state behavior, unstable stacking-fault energies, and adatom migration on the (111) surface, while carrying negative weights for point-defect formation, grain-boundary energies, and adatom migration on the (100) surface. 
In contrast, melting points, phonon DOS, and embedded-dimer tests exhibit consistently low weights across all components. 

This distribution suggests a partial separation between defect-related metrics—both local and extended—and more geometry-independent quantities such as elastic and thermodynamic properties. 
Interestingly, related defect metrics occasionally appear with opposite signs: for example, the migration of (111) adatoms and bulk monovacancies contribute oppositely to the same component, indicating that distinct migration mechanisms are captured differently by various potential families.

\medskip
\noindent\textbf{Component correlations and trade-offs.}  
Pairwise comparisons of potential performance along the three components are shown in Figure~\ref{fig:pca plots}. 
Components~1 and~2 are strongly correlated, producing a negative-diagonal trend: potentials that perform well on one tend to perform well on the other. 
In contrast, comparisons involving Component~3 reveal clear trade-offs, with emerging Pareto fronts along the $+x/-y$ diagonal. 
Potentials that achieve low errors in Components~1 and~2—dominated by equilibrium and formation energies—often perform less accurately for Component~3, which reflects migration and short-range interactions. 

\medskip
\noindent\textbf{Performance by formalism.}  
SNAP machine-learning potentials occupy the lowest-error region of the PCA space and define the leading edge of these Pareto fronts, highlighting their ability to balance accuracy across diverse metrics. 
However, several physically motivated models, including Mishin’s ADP potential and selected EAM and MEAM variants (notably Potentials~1 and~16, both parameterized for Ni–H), also cluster near this frontier. 
Simpler pairwise and Morse-type potentials perform poorly across all components, underscoring their limited transferability. 
Overall, the PCA reveals systematic accuracy trade-offs between equilibrium and defect-related properties and suggests that even mature formalisms such as EAM retain headroom for optimization when applied to complex defect and interface behavior.

\subsection{Vacancy Cluster Coalescence}

Figure~\ref{fig:vacancy clust} summarizes the predicted stability of structured and amorphous vacancy clusters across the tested potentials. 
Because the coalescence of vacancies into larger metastructures remains an active topic of high-accuracy simulation, these results are presented without direct DFT benchmarks. 

Several general trends emerge. 
Most potentials consistently favor either SFTs or amorphous voids across all cluster sizes, while a smaller subset predicts a transition from SFTs at small cluster sizes to voids at larger ones. 
Only one potential exhibits the reverse behavior, forming a void at an intermediate size and reverting to an SFT at the next larger cluster. 
When such transitions occur, they almost always proceed from SFTs to voids with increasing vacancy number. 
Importantly, no clear pattern appears within the same potential formalism: EAM, MEAM, and SNAP models all display diverse preferences, indicating that this behavior is largely independent of functional form and likely sensitive to individual parameterization choices rather than formalism type.

\section{Conclusions}

Benchmarking 34 Ni--Ni interatomic potentials against DFT reference data highlights systematic differences in how formalisms and parameterizations capture microscopic mechanisms in fcc Ni. 
Most models reproduce common benchmarks such as lattice parameters, elastic constants, and surface energies—with good accuracy. However, performance degrades for less common scenarios involving complex interfaces, defect migration, and short-range interactions. 
PCA reveals distinct groupings between equilibrium and defect-related properties and exposes accuracy trade-offs between these categories in the form of Pareto fronts. 
Machine-learning SNAP potentials consistently occupy the low-error frontier, but several EAM, MEAM, and ADP models approach comparable accuracy at a much lower compute cost. 
Future work will extend this benchmarking framework to include real-time validation during machine-learning potential training, enabling more systematic exploration of the accuracy–transferability-cost landscape for transition-metal systems.

\section{Code Availability}

All scripts and configuration files used to generate the results presented here are publicly available at 
\texttt{https://gitlab.com/mattmtl/ni-pots-tests} under the BSD 3-Clause license.

\section{Acknowledgments}
This work was funded by the University Network For Excellence in Nuclear Engineering (UNENE), Mitacs, and the Natural Sciences and Engineering Research Council of Canada (NSERC). The authors are grateful to the Digital Research Alliance of Canada (DRAC), formerly known as Compute Canada, for generous allocation of computer resources.

\nocite{*}

\bibliography{refs}

\appendix

\section{DFT Benchmark Parameters}
\label{appendix:benchmark}

This section details the DFT parameterizations for each benchmark against which potential results are judged.

\medskip
\noindent\textbf{Adsorbed defect behaviour.}
Formation and migration energies of an adatom on a 111 surface is taken from Nakao \emph{et~al.} \cite{doi:10.1021/acs.jpcc.6b03440}
As detailed in their supporting information, the model system is a 3-layer slab followed by a 10~{\AA} void; calculations were performed both with a $3 \times 3$ and $6 \times 6$ supercell.
All energies were calculated in the Vienna \textit{Ab-initio} Simulator Package (VASP) using the Perdew-Burke-Ernzerhoff formulation of the generalized gradient approximation (GGA--PBE) exchange-correlation functional and a projector-augmented wave (PAW) pseudopotential.
The basis set cutoff was set to 400 eV, with K-point grids of $4 \times 4 \times 2$ and $2 \times 2 \times 1$ for each supercell respectively.

Migration energies of an adatom on a 100 surface is taken from Chang \emph{et~al.} \cite{C_M_Chang_2001}
Model systems are a 5-layer slab with a 10~{\AA} void.
Benchmarks are specifically taken from the $5 \times 5$ supercell with a $2 \times 2$ K-point grid; vertical density is not stated.
These figures are calculated using VASP and a GGA--PBE exchange-correlation functional and an ultrasoft pseudopotential.
Basis set cutoff is set to 17.9~Ry.

\medskip
\noindent\textbf{Elastic constants.}
Finite temperature elastic constant benchmarks are taken from Hachet \emph{et~al.} \cite{HACHET2018280}
Calculations were performed using Quantum ESPRESSO's PWSCF code using a GGA--PBE exchange correlation functional and a PAW pseudopotential.
A $16 \times 16 \times 16$ K-point grid was used to sample the primitive cell with a basis set cutoff of 816~eV (approximately 60~Ry).

\medskip
\noindent\textbf{Equation of state.}
Equation of state data is taken from the Materials Cloud Standard Solid-state Pseudopotentials (SSSP) dataset prepared by Prandini et al, using the Vanderbilt 1.2 USPP on, presumably, a primitive cell (1 atom). \cite{prandini_2021_69mvv-cv520}
K-point density is $6 \times 6 \times 6$ with a basis set cutoff of 200~Ry and Marzari-Vanderbilt smearing of 0.02~Ry.

\medskip
\noindent\textbf{Free surface and grain boundary formation.}
Both free surface and grain boundary formation energies are taken from the Materials Virtual Lab Crystalium dataset. \cite{Tran2016, zheng2019grainboundarypropertieselemental}
For free surfaces, the model system contains a $ > $~10~{\AA} slab with a 10~{\AA} void.
Calculations were performed in VASP with a GGA--PBE exchange-correlation functional and a PAW pseudopotential.
Basis set cutoff was set to 400~eV (approximately 29.3~Ry) with a primitive K-point sampling resolution of $50 \times 50 \times 1$ scaled down appropriately in the $x$ and $y$ directions to match the overall cell size.
Grain boundaries required a 25~{\AA} cell normal to the grain boundary plane.
Simulation environments were similar to those in free surfaces, except that the K-point grid densities were 30~/~{\AA} and 45~/~{\AA} for each direction orthogonal to the grain boundary plane.

\medskip
\noindent\textbf{$\gamma$--surface.}
Generalized stacking fault energies ($\gamma$--surfaces) are taken from Hasan \emph{et~al.} \cite{HASAN2020100555}
The model system consisted of 2 slabs at least 36 atoms thick.
Calculations were performed in the CASTEP code with its ultrasoft pseudopotentials and a GGA--PBE exchange-correlation functional.
Reported K-point spacing was 0.08~/~{\AA} in each dimension, with a basis set cutoff of 500~eV (approximately 36.7~Ry).

\medskip
\noindent\textbf{Melting point.}
An \textit{ab-initio} estimate of the melting point is taken from Kim \emph{et~al.} as an extension from a calculated $C_{11}$. \cite{KIM2009254}
DFT calculations were performed in VASP with a GGA--PBE exchange-correlation functional and a PAW pseudopotential.
The model system was a 32-atom supercell with a $6 \times 6 \times 6$ K-point sampling grid.
Basis set cutoff was set to 350~eV (approximately 25.7~Ry).

\medskip
\noindent\textbf{Phonon density of states.}
Phonon DOS calculations are taken from Bhardwaj \emph{et~al.} \cite{bhardwaj2023latticedynamicsrelatedproperties}
As opposed to most other benchmarks, DOS curves are based on all-electron DFT+U calculations of the interatomic force constants from WIEN2K and thus do not use a pseudopotential; the exchange-correlation functional is standard GGA--PBE.
These calculations are done on a $2 \times 2 \times 2 supercell$ sampled with a $5 \times 5 \times 5$ K-point grid.
DOS curves are obtained via the PhonoPy package.

\medskip
\noindent\textbf{Point defect behaviour.}
All point defect formation energies aside from the 110 and 111 dumbbells are taken from Conn\'{e}table \emph{et~al.} \cite{CONNETABLE201577}
Calculations were performed on 108-atom supercells in VASP using the Perdew-Wang GGA exchange-correlation functional (PW91) and a PAW pseudopotential.
Basis set cutoff was set to 600~eV (approximately 44~Ry) and a K-point sampling resolution of $24 \times 24 \times 24$ per 4-atom unit cell, scaled down for supercells to maintain density.
110 and 111 dumbbell formation energies along with all defect migration energy benchmarks are taken from Toijer et al, with calculations also performed in VASP, but using a GGA--PBE exchange-correlation functional. \cite{PhysRevMaterials.5.013602}
256-atom supercells were used, sampled with a $3 \times 3 \times 3$ K-point grid.
Basis set cutoff was set to 350~eV (25.7~Ry).
Migration energies were calculated with the climbing-image nudged elastic band (CI-NEB) method using 3 images. 

\medskip
\noindent\textbf{Embedded dimer.}
Embedded dimer benchmarks are taken from B\'{e}land \emph{et~al.} \cite{BELAND201711}
Calculations were performed on a 32-atom supercell using Quantum ESPRESSO's PWSCF code with a GGA--PBE exchange-correlation functional and an ultrasoft pseudopotential.
Basis set cutoff was set to 104 Ry with an $8 \times 8 \times 8$ K-point sampling grid.

\medskip
\noindent\textbf{Step edge formation.}
Step-edge data is taken from Zhang \emph{et~al.} \cite{doi:10.7566/JPSJ.82.074709}
Calculations were performed using GPAW with a GGA--PBE exchange-correlation functional and a PAW pseudopotential.
The model system is a 17-atom slab delimited by a 15~{\AA} void, sampled with a $5 \times 10 \times 1$ K-point sampling grid.

\end{document}